\documentclass[12pt,twoside,english,preprint]{article}
\usepackage{mathptmx}

\usepackage[T1]{fontenc}
\usepackage[latin9]{inputenc}
\usepackage{geometry}
\usepackage{amstext}
\usepackage{setspace}
\usepackage{esint}
\makeatletter
\@ifundefined{date}{}{\date{}}
\usepackage{tikz}  
\usepackage{float} 
\usepackage{pgfplots}  
 \usetikzlibrary{arrows}

\makeatother

\usepackage{babel}
\begin{document}
\title{\noindent Physical states and correction terms of the Supersymmetric
$c=1$ Model}
\author{O. El Deeb}
\maketitle
\begin{center}
Email: omar.eldeeb@lau.edu.lb
\par\end{center}

\begin{center}
Address: Department of Natural Sciences
\par\end{center}

\begin{center}
Lebanese American University, Beirut, Lebanon
\par\end{center}
\begin{abstract}
In this article, we investigate the supersymmetric $c=1$ model of
superstring theory and demonstrate how the spectrum of states is expanded
and new symmetries of the theory are generated by the existence of
ghost cohomologies. As a result, we establish significant connections
between two-dimensional supergravity and physical theories in higher
dimensions. Additionally, we provide a comprehensive guide for constructing
$BRST$-invariant and nontrivial vertex operators and carry out explicit
computations to determine the correction terms needed to maintain
$BRST$ invariance of the corresponding currents.
\end{abstract}
keywords: Supersymmetry, Supergravity, Ghost Cohomology, String theory,
BRST invariance

\section{Introduction}

The formalism of ghost cohomologies is an approach used to study the
non-perturbative dynamics of strings by exploring the important information
contained in physical operators that are $BRST$ (Becchi-Rouet-Stora-Tyutin)
invariant but not always manifestly gauge-invariant. This approach
allows researchers to investigate the behavior of strings beyond the
perturbative regime and gain a better understanding of the non-perturbative
physics of gauge theories. In particular, it is used to study the
properties of special vertex operators and their relation to non-perturbative
strings. These vertex operators are not always manifestly $BRST$-invariant,
meaning they do not commute with the supercurrent terms of the $BRST$
charge. By adding $b-c$ ghost dependent terms to the expressions
of these operators, researchers can restore their $BRST$ invariance
and better understand their properties. In string theory, the $b-c$
ghosts are auxiliary fields that are introduced to make the worldsheet
theory invariant under the $BRST$ symmetry, which is a symmetry that
plays a crucial role in the quantization of gauge theories. The $b$
and $c$ ghosts are worldsheet fields that satisfy a set of algebraic
relations and can be thought of as anticommuting fields. The $b-c$
ghost dependent terms are terms that involve both the $b$ and $c$
ghosts and are added to the expressions of certain vertex operators
in string theory. The addition of $b-c$ ghost dependent terms allows
us to better understand the properties of these vertex operators and
gain insights into the non-perturbative dynamics of strings. Overall,
the formalism of ghost cohomologies provides a powerful tool for exploring
the non-perturbative dynamics of strings and gaining insights into
the physics of gauge theories.

In the first section, we review the concepts of $BRST$ quantization
and vertex operator formalism in the perturbative Ramond-Neveu-Schwarz
($RNS$) superstring theory \cite{GSW,POL,Non-Abelian,Adam}. We also
examine global space-time symmetries and the generator of this symmetry,
which is given by a special type of the worldsheet current violating
the principle of ghost picture equivalence \cite{POLY2}. The perturbative
$RNS$ superstring theory is a type of string theory that describes
the behavior of fundamental strings in terms of the motion of their
worldsheet. The theory incorporates both bosonic and fermionic fields
on the worldsheet and has supersymmetry, which allows for the cancellation
of divergences that arise in the theory. In the $RNS$ formalism,
the worldsheet fields include the spacetime coordinates of the string,
along with a set of fermionic fields known as spinors, which describe
the fermionic degrees of freedom of the string. The theory has a set
of symmetries, including Lorentz invariance and worldsheet reparametrization
invariance, which constrain the dynamics of the theory. The perturbative
RNS superstring theory is an important framework for studying the
properties of strings and their interactions. It provides a foundation
for the development of more advanced string theories, such as the
superstring theories and $M$-theory, which incorporate additional
symmetries and degrees of freedom. 

We explore the properties of special vertex operators and their relation
to non-perturbative strings, using examples of critical string theory
and the supersymmetric $c=1$ model \cite{LiTs,KlPo,Witten}. the
supersymmetric $c=1$ model is a two-dimensional quantum field theory
that describes the behavior of strings in a specific spacetime background.
The theory incorporates both bosonic and fermionic fields on the worldsheet,
and has supersymmetry, which allows for the cancellation of divergences
that arise in the theory. The $c=1$ model has a central charge of
one, which means that it has only one massless mode and is critical,
meaning that it has no tachyons and is free of divergences. The theory
has a set of symmetries, including Lorentz invariance and worldsheet
reparametrization invariance, which constrain the dynamics of the
theory. The supersymmetric $c=1$ model is an important framework
for studying the properties of strings and their interactions in a
specific spacetime background. It has applications in areas such as
condensed matter physics, where it can be used to model the behavior
of certain two-dimensional materials. The study of this model has
also led to insights into more general aspects of string theory and
its connections to other areas of physics \cite{Vafa1,Vafa2,Pelliconi,Montero}.

We introduce the notion of ghost cohomologies and discuss their appearance
based on the approach used in \cite{POLY1}. Section 2.3 explores
the question of $BRST$ invariance of vertex operators from the ghost
cohomologies of positive and negative ghost numbers \cite{POLY3-1}.
An interesting property of these vertex operators is that they are
not manifestly $BRST$-invariant as they don\textquoteright t commute
with the supercurrent terms of the $BRST$ charge \cite{Poly2019,SuperString,Superstring2,2D}.
The main importance of the general prescription that we provide here
is that it allows to restore their $BRST$ invariance by adding $b-c$
ghost dependent terms to the expressions of these operators and demonstrate
how this strategy generally works on the example of the ghost-matter
mixing five-form state of the critical $NSR$ superstring theory in
ten dimensions. More importantly, we provide an explicit calculation
of the correction terms restoring the $BRST$ invariance of the $T_{1,2}$
current. In the concluding section, we summarize our results and discuss
some possible implications of our results and suggest directions for
future work.

\section{The $RNS$ Superstring Theory}

String theory has been always considered as a promising model that
might be finally able to unify the fundamental interactions of nature
and it is also one of the best candidates for the construction of
a consistent theory of quantum gravity. While the perturbative theory
of strings seems to be already well explored, we still lack a complete
and adequate string-theoretic formalism in the strongly coupled non-perturbative
regimes. Strings were initially introduced in an attempt to find a
solution for the problem of quark confinement as it is well-known
that quarks exist in the bound state only as the interaction force
between them grows with increasing distance \cite{POLYAM}. Up until
now, string theory has not been able to fully explain the quark confinement
mechanism because it turned out to be hard to find an open string
model whose partition function can exactly reproduce the expectation
value of the Wilson\textquoteright s loop in Quantum Chromodynamics
($QCD$)\cite{POLYAM}. Nevertheless, several significant acheivements
have been made in the string-theoretic approach to $QCD$ in the recent
years \cite{MALCA,Witt,Gubser}. AdS/CFT correspondence between string
energies on the $AdS_{5}\times S^{5}$ background with the anomalous
dimensions of gauge invariant operators in supersymmetric four-dimensional
$SU(N)$ gauge theory, with several applications and extensions, has
been successfully developed and studied \cite{ArtFro,Konishi,5loop,eldeeb,BelSpod,kallosh}.
Possible applications of string theory go far beyond $QCD$ and the
theory of strong interaction as well. This is why it is strongly beleived
that string theory appears to be a candidate for a model unifying
gravity with the standard model.

String theory is a promising model that has the potential to unify
fundamental interactions of nature and provide a consistent theory
of quantum gravity. While the perturbative theory of strings has been
extensively studied, a complete and appropriate string-theoretic formalism
is still lacking in strongly coupled non-perturbative regimes. Originally,
strings were proposed as a solution to the problem of quark confinement,
which arises because quarks exist only in bound states as their interaction
force increases with distance \cite{POLYAM}. However, string theory
has not yet fully explained the quark confinement mechanism due to
the difficulty in finding an open string model that can accurately
reproduce the expectation value of the Wilson's loop in Quantum Chromodynamics
($QCD$). Despite this, significant progress has been made in the
string-theoretic approach to $QCD$ in recent years \cite{MALCA,Witt,Gubser},
particularly through the $AdS/CFT$ correspondence between string
energies on the $AdS_{5}\times S^{5}$ background and the anomalous
dimensions of gauge invariant operators in supersymmetric four-dimensional
$SU(N)$ gauge theory \cite{ArtFro,Konishi,5loop,eldeeb,eldeeb2,BelSpod,kallosh}.
This correspondence has numerous applications and extensions, and
the potential applications of string theory extend beyond $QCD$ and
the theory of strong interactions. Therefore, string theory is considered
a promising candidate for a model that unifies gravity with the standard
model \cite{Unification1,Unification2,Unification3}.

First, we provide an overview of the fundamental concepts of string
dynamics before extending them to the supersymmetric scenario. We
use local coordinates,$\xi_{1}\text{ and }\xi_{2}$, to parameterize
the worldsheet, and $X_{m}\ (m=0,1,...,d-1)$ to denote the $d$-dimensional
space-time coordinates. The string action is defined as follows:

\begin{equation}
S_{string}=-\frac{1}{4\pi}\int d^{2}z\sqrt{\gamma}\gamma^{ab}\partial_{a}X^{m}\partial_{b}X^{n}(\xi_{1},\xi_{2})\eta_{mn}\label{eq:string action}
\end{equation}

where $\gamma^{ab}(\xi_{1},\xi_{2})(a,b=1,2)$ is the induced worldsheet
metric and $\eta^{mn}$ is the Minkowski metric \cite{PolyAM1}. In
addition, the action is symmetric under the reparametrizations of
the local worldsheet coordinates:

\begin{equation}
\begin{array}{cc}
\xi_{1}\rightarrow f_{1}=f_{1}(\xi_{1},\xi_{2})\\
\xi_{2}\rightarrow f_{2}=f_{2}(\xi_{1},\xi_{2})
\end{array}\label{eq:local worldsheet coordinates}
\end{equation}

The reparametrization symmetry (2) is a crucial local bosonic gauge
symmetry in the framework of string theory. By making use of these
gauge transformations, the worldsheet metric $\gamma^{ab}$ can be
transformed into the conformally flat form $\gamma^{ab}\rightarrow e^{\varphi}(\xi_{1},\xi_{2})\delta^{ab}$,
where $\varphi$ is a function of the worldsheet coordinates. Additionally,
the action (1) is invariant under Weyl rescalings of the metric, $\gamma^{ab}\rightarrow e^{\sigma}(\xi_{1},\xi_{2})\gamma^{ab}$.
It is possible to eliminate the scale factor of $e^{\varphi}$ in
the worldsheet metric, which reduces action (1) to that with a flat
metric. When subject to reparametrizations and Weyl rescalings, the
integration measure undergoes the following transformations \cite{PolyAM1}:

\begin{equation}
D[\gamma^{ab}]\rightarrow D[\varphi]e^{-S_{Liouville}}D[b]D[c]e^{-S_{b-c}}
\end{equation}

The Faddeev-Popov determinant, which is similar to fixing the conformal
gauge, arises from the action of the fermionic b-c reparametrization
ghost fields $S_{bc}=\frac{1}{4\pi}\int d^{2}z(b\bar{\partial}c+\bar{b}\partial\bar{c})$
\cite{GSW,PolyAM1,FaPo}. To make the calculation more convenient,
the b and c ghost fields are bosonized using a single free bosonic
field according to \cite{FMS} with $b=e^{-\sigma}$ and $c=e^{\sigma}$.
The Liouville field $\varphi$ has an action given by 
\begin{equation}
S_{Liouville}=\frac{D-26}{36\pi}\int d^{z}z(\partial\varphi\bar{\partial}\varphi+2\mu_{0}be^{b\varphi})
\end{equation}
which reflects the anomaly of the functional integration measure $D[\varphi]$
under the Weyl rescaling. The cosmological constant constant is denoted
by $\mu_{0}$, and $b$ is a constant determined by the Liouville
field's background charge $Q=b+\frac{1}{b}=\sqrt{\frac{25-d}{3}}$
to make the total central charge of the system zero $c_{X}+c_{b-c}+c_{\varphi}=0$
\cite{GSW,POL,PolyAM1}. The central charge in conformal field theory
($CFT$) is determined from the two-point correlation functions of
the stress-energy tensors of the appropriate fields \cite{BPZ,Ginsparg},
such that 
\begin{equation}
<T(z)T(w)>=\frac{\frac{c}{2}}{(z-w)^{4}}
\end{equation}
where $T$ is defined as $T_{zz}\text{ and }T_{ab}=2\pi(\gamma)^{\frac{1}{2}}\frac{\delta S}{\delta\gamma^{ab}}$. 

By using the expressions for the stress-tensors of $X,\varphi$, and
the ghost fields, the central charge can be calculated.

\begin{equation}
\begin{array}{cc}
T_{x}=-\frac{1}{2}\partial X_{m}\partial X^{m}\\
T_{\varphi}=-\frac{1}{2}(\partial\varphi)^{2}+\frac{Q}{2}\partial^{2}\varphi\\
T_{bc}=\frac{1}{2}(\partial\sigma)^{2}+\frac{3}{2}\partial^{2}\sigma
\end{array}
\end{equation}

together with the operator products:

\begin{equation}
\begin{array}{cc}
X^{m}(z)X^{n}(w)\sim-\eta^{mn}ln(z-w)\\
\varphi(z)\varphi(w)\sim-ln(z-w)\\
\sigma(z)\sigma(w)\sim ln(z-w)
\end{array}
\end{equation}

it is possible to show that $c_{X}=d,\ c_{\varphi}=1+3Q^{2},\ c_{b-c}=-26.$
The full matter-ghost action is then given by

\begin{equation}
S=\frac{-1}{4\pi}\int d^{2}z\partial X_{m}\partial X^{m}+S_{Liouville}+S_{b-c}
\end{equation}

The $RNS$ model incorporates two additional anticommuting Grassmann
coordinates $\theta$ and $\bar{\theta}$ that extend the worldsheet
coordinates ($z,$ $\bar{z}$). These Grassmann coordinates are defined
in such a way that they satisfy certain conditions \cite{WessBA}:

\begin{equation}
\begin{array}{cc}
\theta^{2}=\bar{\theta}^{2}=0, & \bar{\theta}\theta=-\theta\bar{\theta}\\
\int d\theta=0, & \int d\theta\theta=1
\end{array}
\end{equation}

The worldsheet integration is replaced by the superspace integral
$\int d^{2}z\rightarrow\int d^{2}zd^{2}\theta$ while concurrently
replacing the derivatives in $z$ and $\bar{z}$ by their covariant
counterparts:

\begin{equation}
\begin{array}{cc}
\partial_{z}\rightarrow D_{z}=\partial_{\theta}+\theta\partial_{z}\text{ , } & \partial_{\bar{z}}\rightarrow D_{\bar{z}}=\partial_{\theta}+\bar{\theta}\partial_{\bar{z}}\end{array}
\end{equation}

The expansions of the superfields are given by:

\begin{equation}
\begin{array}{cc}
X^{m}(z,\bar{z},\theta,\bar{\theta})=X^{m}(z,\bar{z})+\theta\psi^{m}(z,\bar{z})+\bar{\theta}\bar{\psi}^{m}(z,\bar{z})+\theta\bar{\theta}H^{m}(z,\bar{z})\\
\varphi(z,\bar{z},\theta,\bar{\theta})=\varphi(z,\bar{z})+\theta\lambda(z,\bar{z})+\bar{\theta}\bar{\lambda}(z,\bar{z})+\theta\bar{\theta}F(z,\bar{z})\\
C(z,\theta)=c(z)+\theta\gamma(z)\\
\bar{C}(\bar{z},\bar{\theta})=\bar{c}(\bar{z})+\bar{\theta}\bar{\gamma}(\bar{z})\\
B(z,\theta)=\beta(z)+\theta b(z)\\
\bar{B}(\bar{z},\bar{\theta})=\bar{\beta}(\bar{z})+\bar{\theta}\bar{b}(\bar{z})
\end{array}
\end{equation}

Integrating out $\theta$ and $\bar{\theta}$, it can be shown that
the full ghost-matter action of the $RNS$ superstring theory in the
superconformal gauge is \cite{DKKMMSS}:

\begin{equation}
\begin{array}{cc}
S_{RNS}=-\frac{1}{4\pi}\int d^{2}z(\partial X_{m}\bar{\partial}X^{m}+\psi_{m}\bar{\partial}\psi^{m}+\bar{\psi}_{m}\partial\bar{\psi}^{m})+S_{ghost}+S_{Liouville}\\
S_{ghost}=\frac{1}{2\pi}\int d^{2}z(b\bar{\partial}c+\bar{b}\partial\bar{c}+\beta\bar{\partial}\gamma+\bar{\beta}\partial\bar{\gamma})\\
S_{Liouville}=\frac{d-10}{36\pi}\int d^{2}z(\partial\varphi\bar{\partial}\bar{\varphi}+\lambda\bar{\partial}\lambda+\bar{\lambda}\partial\bar{\lambda}-F^{2}+2\mu_{0}be^{b\varphi}(ib\lambda\bar{\lambda}-F))\\
Q=b+\frac{1}{b}=\sqrt{\frac{9-d}{2}}
\end{array}
\end{equation}

We bosonize the superconformal $\beta$ and $\gamma$ ghosts in terms
of the pair of free $2d$ scalar bosons $\phi$ and $\chi$ . The
bosonization relations are given by \cite{FMS}:

\begin{equation}
\begin{array}{cc}
\gamma=e^{\phi-\chi}, & \beta=e^{\chi-\phi}\partial_{\chi}\\
<\chi(z)\chi(w)> & =-<\phi(z)\phi(w)>=ln(z-w)
\end{array}
\end{equation}

and the full matter$+$ghost stress-energy tensor is

\begin{equation}
\begin{array}{cc}
T_{matter}=-\frac{1}{2}\partial X_{m}\partial X^{m}-\frac{1}{2}\partial\psi_{m}\psi^{m}\\
T_{ghost}=\frac{1}{2}(\partial\sigma)^{2}+\frac{3}{2}\partial^{2}\sigma+\frac{1}{2}(\partial\chi)^{2}+\frac{1}{2}\partial^{2}\chi-\frac{1}{2}(\partial\phi)^{2}-\partial^{2}\phi\\
T_{Liouville}=-\frac{1}{2}(\partial\varphi)^{2}+\frac{Q}{2}\partial^{2}\varphi
\end{array}
\end{equation}

\paragraph{Vertex operators}

In the string theory framework presented in this section, the oscillation
modes are typically interpreted as fundamental particles, solitons,
black holes, or $D$-branes. These entities are commonly described
by vertex operators \cite{GSW,POL,BRST} with the following form:

\begin{equation}
V=P(\partial X^{m},\partial^{2}X^{m},...,\psi^{m},\partial\psi^{m},ghosts...)e^{ik_{m}X^{m}}
\end{equation}

Here, $P$ is a polynomial in the fields and their derivatives, and
$k_{m}$ corresponds to the momentum. A vertex operator is deemed
physical if it belongs to the $BRST$ cohomology. The noteworthy characteristic
of all gauge symmetries in such theories is that any local gauge symmetry
of the theory automatically encompasses its invariance under another
type of symmetry transformations, referred to as $BRST$ symmetry
\cite{BRST,FaSl}. This implies that if an action of the theory is
locally invariant under a certain local gauge symmetry, it would also
be invariant under transformations of the same form as the gauge transformations,
but with the local gauge parameter replaced by the corresponding Faddeev-Popov
ghost of the opposite statistics. For the $RNS$ model, the $BRST$
charge expression is given precisely as:

\begin{equation}
Q_{BRST}=\oint\frac{dz}{2i\pi}\left\{ cT-bc\partial c-\frac{1}{2}\gamma\psi_{m}\partial X^{m}-\frac{1}{4}\gamma^{2}b\right\} 
\end{equation}

To be considered physical vertex operators in the RNS model, they
must be invariant under BRST transformations, meaning that they satisfy
${Q_{BRST},V}=0$. Operators that are $BRST$-exact and can be written
as $V={Q_{brst},W}$ should be excluded. Open strings have an open
line configuration, while closed strings have a loop configuration,
with the simplest topology being a sphere. To be physical, these operators
must be primary fields of dimension 1 for open strings and $(1,1)$
for closed strings. In Conformal Field Theory ($CFT$), primary fields
of dimension ($h$, $\bar{h}$) are observables $\varphi^{h,\bar{h}}$
that transform under conformal transformations as $z\rightarrow f(z);\bar{z}\rightarrow f(\bar{z})$
according to as 
\begin{equation}
\varphi^{h,\bar{h}}(z,\bar{z})\rightarrow\left(\frac{df}{dz}\right)^{h}\left(\frac{d\bar{f}}{d\bar{z}}\right)^{\bar{h}}\left(f(z),\bar{f}(\bar{z})\right)
\end{equation}

Therefore, the operator product expansion (OPE) of the stress-energy
tensor has a simple form:

\begin{equation}
T(z)\varphi^{h,\bar{h}}(w,\bar{w})=\frac{h\varphi^{h,\bar{h}}(w,\bar{w})}{(z-w)^{2}}+\frac{\partial\varphi^{h,\bar{h}}(w,\bar{w})}{z-w}+O(z-w)^{0}
\end{equation}

\section{Discrete States and Currents}

\subsection{Ghost Cohomologies in $RNS$ Model}

Physical states in the $RNS$ superstring theory are vertex operators
that are both nontrivial and $BRST$-invariant. These operators can
be defined up to transformations by the picture-changing operator
$\Gamma=[Q_{brst},\xi]$ and its inverse operator $\Gamma^{-1}=c\partial\xi e^{-2\phi}$,
where $\xi=e^{\chi}$ and $\phi,\chi$ is the pair of the bosonized
superconformal ghosts, which consists of the product of the $BRST$
charge and the pair of bosonized superconformal ghosts. The inverse
of the picture-changing operator increments or decrements the ghost
number of the operator by $1$. Thus, each string excitation can be
described by an infinite set of physically equivalent operators. 

\begin{equation}
\begin{array}{c}
:\Gamma V^{(n)}:=V^{(n+1)}+\{Q_{brst},...\}\\
:\Gamma^{-1}V^{(n)}:=V^{(n-1)}+\{Q_{brst},...\}
\end{array}
\end{equation}

The location of picture-changing operators inside correlation functions
can vary, resulting in a full derivative in the supermoduli space.
This ensures their picture invariance under appropriate moduli integration.
However, global singularities arise if the correlation function contains
vertex operators that diverge faster than $(z-z_{n})^{-2}$, where
$z_{n}$ is the insertion point on the worldsheet \cite{POLY3}. If
such an operator is present, the moduli integration of the full derivative
term would result in a nonzero boundary contribution, and the correlation
function would be picture-dependent. An example of such a vertex operator
in the critical $RNS$ superstring theory is the five-form state:

\begin{equation}
V_{5}(k)=H(k)\oint\frac{dz}{2i\pi}U_{5}(z)\equiv H_{m_{1}...m_{5}}(k)\oint\frac{dz}{2i\pi}e^{-3\phi}\psi^{m_{1}}...\psi^{m_{5}}e^{ikX}(z)
\end{equation}
This operator exists at all the negative pictures below $-3$ but
has no versions at pictures $-2$ or higher \cite{POLY2}. Since $V_{5}$
is annihilated by $\Gamma$, we have

\begin{equation}
0\equiv\lim_{w\rightarrow z}\Gamma^{-1}(u)\Gamma(w)V_{5}(z)=V_{5}(z)+\lim_{w\rightarrow z}\{Q_{brst},\Lambda(u,w)V_{5}(z)\}
\end{equation}
hence $V_{5}$ is the $BRST$ commutator $V_{5}(z)=-\lim_{w\rightarrow z}\left\{ Q_{brst},\Lambda(u,w)V_{5}(z)\right\} .$
It can be written as

\begin{equation}
\begin{array}{cc}
\lim_{w\rightarrow z} & \left\{ Q_{brst},\Lambda(u,w)V_{5}(z)\right\} =\lim_{w\rightarrow z}\left\{ Q_{brst},\Gamma^{-1}(u)\left(\xi(w)-\xi(u)\right)V_{5}(z)\right\} \\
 & =\lim_{w\rightarrow z}\Gamma^{-1}(u)\left(\Gamma(w)-\Gamma(u)\right)V_{5}(z)
\end{array}
\end{equation}
and the commutator is given by $V_{5}(z)=\left\{ Q_{brst},\xi\Gamma^{-1}(u)V_{5}(z)\right\} .$
Since the derivatives of $\Gamma$ are all $BRST$-trivial: $\partial^{n}\Gamma=\{Q_{brst},\partial^{n}\xi\};\ n=1,2,...$
, and one can write

\begin{equation}
\Gamma(w)=\Gamma(z)+\left\{ Q_{brst},\sum_{n}\frac{(w-z)^{n}}{n!}\partial^{n}\xi\right\} 
\end{equation}
The non-singular $OPE$ of $\Gamma$with $U_{5}$ is given by

\begin{equation}
\Gamma(z)U_{5}(w)\sim(z-w)^{2}e^{-2\phi}\psi^{m_{1}}...\psi^{m_{4}}(i(k\psi)\psi^{m_{5}}+\partial X_{m_{5}})e^{ikX}
\end{equation}
Our goal here is to derive the correction terms restoring the $BRST$-invariance
of $W_{5}$ and to demonstrate the nonzero correlator involving the
$V_{5}$ and $W_{5}$ operators. We start with the $BRST$ invariance
restoration and we consider the $BRST$ charge given by

\begin{equation}
Q_{brst}=\oint\frac{dz}{2i\pi}\left(cT-bc\partial c-\frac{1}{2}\gamma\psi_{m}\partial X^{m}-\frac{1}{4}\gamma^{2}b\right)
\end{equation}
Using the operator $L(z)=-4ce^{2\chi-2\phi}\equiv:\xi\Gamma^{-1}:$
we find that it satisfies $\{Q_{brst},L\}=1$. Next, consider a non-invariant
operator $V$ satisfying $\{Q_{brst},V\}=W$, then $W$ is $BRST$-exact,
and the transformation $V\rightarrow V_{inv}=V-LW$ restores $BRST$-invariance.
Applying the same scheme for $W_{5}$, we have

\begin{equation}
[Q_{brst},W_{5}]=H_{m_{1}...m_{5}}(k)\oint\frac{dz}{2i\pi}e^{2\phi-\chi+ikX}R_{1}^{m_{1}...m_{5}}(z)+be^{3\phi-2\chi+ikX}R_{1}^{m_{1}...m_{5}}(z)
\end{equation}

where

\begin{equation}
\begin{array}{cc}
R_{1}^{m_{1}...m_{5}}(z)=-\frac{1}{2}\psi^{m_{1}}...\psi^{m_{5}}(\psi\partial X)-\frac{1}{2}\psi^{[m_{1}...m_{4}}(\partial^{2}X^{m_{5}]}+\partial X^{m_{5}]}(\partial\phi-\partial\chi))\\
-\frac{i}{2}\psi^{m_{1}}...\psi^{m_{5}}(k\psi)(\partial\phi-\partial\chi)+(k\partial\psi)
\end{array}
\end{equation}

and

\begin{equation}
R_{1}^{m_{1}...m_{5}}(z)=-\frac{1}{4}(2\partial\phi-2\partial\chi\partial\sigma)\psi^{m_{1}}...\psi^{m_{5}}
\end{equation}

Evaluating the $OPE$ of $L$ we obtain

\begin{equation}
\begin{array}{c}
W_{5inv}(k,w)=H_{m_{1}...m_{5}}(k)\{\oint\frac{dz}{2i\pi}e^{\phi}\psi^{m_{1}}...\psi^{m_{5}}e^{ikX}\\
-\frac{1}{2}\oint_{w}\frac{dz}{2i\pi}(z-w)^{2}:L\partial_{z}^{2}(e^{2\phi-\chi}R_{1}^{m_{1}...m_{5}}+e^{3\phi-2\chi}R_{2}^{m_{1}...m_{5}}(z)):\}\\
=H_{m_{1}...m_{5}}(k)\{\oint\frac{dz}{2i\pi}e^{\phi}\psi^{m_{1}}...\psi^{m_{5}}e^{ikX}-2\oint_{w}\frac{dz}{2i\pi}(z-w)^{2}ce^{\chi}R_{1}^{m_{1}...m_{5}}(k,z)\}
\end{array}
\end{equation}

There are three types of ghost cohomologies: positive ghost number
cohomology, negative ghost number cohomology, and zero ghost cohomology.
A formal definition of ghost cohomologies was given in \cite{POLY2,POLY1}
and is summarized by
\begin{enumerate}
\item The positive ghost number cohomology, denoted as $H_{N}$ (where $N$
is a positive integer), consists of physical vertex operators that
exist at positive superconformal ghost pictures $n\geq N$ and are
annihilated by the inverse picture-changing operator $\Gamma^{-1}$
at picture $N$. This means that the picture $N$ is the minimal positive
picture at which the operators $V\subset H_{N}$ can exist.
\item The negative ghost number cohomology, denoted as $H_{-N}$ (where
$N$ is a positive integer), consists of physical vertex operators
that exist at negative superconformal pictures $n\leq N$ and are
annihilated by the direct picture-changing at picture $-N$.
\item By definition, zero ghost cohomology, denoted as$H_{0}$, consists
of operators that exist at all pictures. Standard string perturbation
theory involves elements of $H_{0}$. The standard string perturbation
theory thus involves the elements of $H_{0}$. The picture $-3$ and
picture $+1$ five-forms considered above are the elements of $H_{-3}$
and $H_{1}$ respectively.
\item There is a generic isomorphism between positive and negative ghost
cohomologies:$H_{-N-2}\sim H_{N}$; $N\geq1$. This means that to
any element of $H_{-N-2}$, there corresponds an element from $H_{N}$
obtained by replacing $e^{-(N+2)\phi}$ with $e^{N\phi}$ and then
adding the appropriate $b-c$ ghost terms in order to restore the
$BRST$-invariance.
\end{enumerate}
The example of the picture $-3$ and picture $+1$ five-forms discussed
earlier belong to $H_{-3}$ and $H_{1}$, respectively. By understanding
ghost cohomologies, researchers in string theory can better understand
the behavior of strings in different spacetime backgrounds and calculate
various physical quantities such as scattering amplitudes and correlation
functions.

\subsection{Extended Discrete States in $c=1$ Supersymmetric Model}

Non-critical one-dimensional string theory is known to include \textquotedbl discrete
states\textquotedbl{} with non-standard $b-c$ ghost numbers $0$
and $2$ in its physical state spectrum. If we consider the$c=1$
model, which is supersymmetrized on the worldsheet and coupled to
the super Liouville field, the action of the system on the worldsheet
in conformal gauge is given by:

\begin{equation}
\begin{array}{c}
S=S_{\chi-\psi}+S_{L}+S_{b-c}+S_{\beta-\gamma}\\
S_{\chi-\psi}=\frac{1}{4\pi}\int d^{2}z\left\{ \partial X\bar{\partial}X+\psi\bar{\partial}\psi+\bar{\psi}\partial\bar{\psi}\right\} \\
S_{L}=\frac{1}{4\pi}\int d^{2}z\left\{ \partial\varphi\bar{\partial}\varphi+\lambda\bar{\partial}\lambda+\bar{\lambda}\partial\bar{\lambda}-F^{2}+2\mu_{0}be^{b\varphi}(ib\lambda\bar{\lambda}-F)\right\} \\
S_{b-c}+S_{\beta-\gamma}=\frac{1}{4\pi}\int d^{2}z\left\{ b\bar{\partial}c+\bar{b}\partial\bar{c}+\beta\bar{\partial}\gamma+\bar{\beta}\partial\bar{\gamma}\right\} 
\end{array}
\end{equation}
where $Q\equiv b+b^{-1}$ is the background charge. The stress tensors
of the matter and the ghost systems and the standard bosonization
relations for the ghosts are given by

\begin{equation}
\begin{array}{c}
T_{m}=-\frac{1}{2}(\partial X)^{2}-\frac{1}{2}\partial\psi\psi-\frac{1}{2}(\partial\varphi)^{2}+\frac{Q}{2}\partial^{2}\varphi\\
T_{gh}=\frac{1}{2}(\partial\sigma)^{2}+\frac{3}{2}\partial^{2}\sigma+\frac{1}{2}(\partial\chi)^{2}+\frac{1}{2}\partial^{2}\chi-\frac{1}{2}(\partial\phi)^{2}-\partial^{2}\phi\\
c=e^{\sigma}\ ,\ b=e^{-\sigma}\\
\gamma=e^{\phi-\chi}\ ,\ \beta=e^{\chi-\phi}\partial\chi
\end{array}
\end{equation}
 According to the prescription provided by \cite{POLY1}, the $SU(2)$
algebra is generated by the currents:

\begin{equation}
\begin{array}{c}
T_{0,0}=\oint\frac{dz}{2i\pi}\partial X\\
T_{0,1}=\oint\frac{dz}{2i\pi}e^{iX}\psi\\
T_{0,-1}=\oint\frac{dz}{2i\pi}e^{-iX}\psi
\end{array}
\end{equation}
It is important to note that the currents of the form

\begin{equation}
\begin{array}{c}
T_{-n,m}=\oint\frac{dz}{2i\pi}P_{-n,m}(\partial X,\partial^{2}X,...,\psi,\partial\psi,...)e^{-n\phi+imX}\\
|m|\leq n-1
\end{array}
\end{equation}

The Virasoro primary states with negative ghost number $-n$, which
are typically not $BRST$ invariant, are usually annihilated by the
picture-changing operator. The states $P_{-n,m}$ are polynomial expressions
in $\partial X,\ \psi,$ and their derivatives, with a conformal weight
of $h=1/2(n^{2}-m^{2})+n+1$, ensuring that the integrals have a total
dimension of $1$. To begin, one starts with the Liouville-dressed
tachyonic Virasoro primaries for a given $n$

\[
\oint\frac{dz}{2i\pi}V_{l}=\oint\frac{dz}{2i\pi}e^{ilX+(l-1)\varphi}(l\psi-i(l-1)\varphi)
\]
with integer $l$ and acts on them with various combinations of the
lowering $T$-operators. The obtained operators will be the multiplets
of $SU(n)$, including the operators of $BRST$ cohomologies with
non-trivial ghost dependence.

\paragraph{Ghost dependent Discrete States}

We consider here three $SU(2)$ currentstaken at different ghost pictures.
The first example is the generator given by the worldsheet integral

\begin{equation}
T_{-3,2}=\oint\frac{dz}{2i\pi}e^{-3\phi+2iX}\psi(z)
\end{equation}
This operator is annihilated by the picture-changing transformation.
By taking the lowering operator $T_{0,-1}=\oint\frac{dz}{2i\pi}e^{-iX}\psi(z)$
of $SU(2)$, we obtain the following extra five generators in the
ghost number $-3$ cohomology:

\begin{equation}
\begin{array}{c}
T_{-3,2}=\oint\frac{dz}{2i\pi}e^{-3\phi+2iX}\psi(z)\\
T_{-3,1}=\oint\frac{dz}{2i\pi}e^{-3\phi+iX}(\partial\psi\psi+\frac{1}{2}(\partial X)^{2}+\frac{i}{2}\partial^{2}X)(z)\\
T_{-3,-1}=\oint\frac{dz}{2i\pi}e^{-3\phi-iX}(\partial\psi\psi+\frac{1}{2}(\partial X)^{2}-\frac{i}{2}\partial^{2}X)(z)\\
T_{-3,0}=\oint\frac{dz}{2i\pi}e^{-3\phi}(\partial^{2}X\psi-2\partial X\partial\psi)(z)\\
T_{-3,-2}=\oint\frac{dz}{2i\pi}e^{-3\phi-2iX}\psi(z)
\end{array}
\end{equation}

The following step involves demonstrating that when combined with
the three standard $SU(2)$ generators $T_{0,0}$, $T_{0,1}$, and
$T_{0,-1}$, the operators form eight generators of $SU(3)$, where
$T_{0,0}$ and $T_{-3,0}$ create the Cartan subalgebra of $SU(3)$.
The computation of the commutators of some of the generators can be
complicated. However, it can be simplified by noting that the operators
$T_{-n,m}$ of ghost cohomology $-n$ (where$n=3,4,...$) are equivalent
to operators from the positive ghost number $n-2$ cohomologies, up
to certain $b-c$ ghost terms necessary for maintaining the $BRST$-invariance
of the operators with positive superconformal ghost numbers. To simplify
the process, it is convenient to redefine the operators

\begin{equation}
\begin{array}{c}
L=\frac{i}{2}T_{0,0},\ H=\frac{i}{3\sqrt{2}}T_{-3,0}\\
G_{+}=\frac{1}{2\sqrt{2}}(\sqrt{2}T_{0,1}+T_{-3,1}),\ G_{-}=\frac{1}{2\sqrt{2}}(\sqrt{2}T_{0,1}-T_{-3,1})\\
F_{+}=\frac{1}{2\sqrt{2}}(\sqrt{2}T_{0,-1}+T_{-3,-1})\\
F_{-}=\frac{1}{2\sqrt{2}}(\sqrt{2}T_{0,-1}-T_{-3,-1})\\
G_{3}=\frac{1}{\sqrt{2}}T_{-3,2},\ F_{3}=\frac{1}{\sqrt{2}}T_{-3,-2}
\end{array}
\end{equation}

Consequently, the commutators of the operators $L$ and $H$ are given
by

\begin{equation}
\begin{array}{c}
[L,H]=0\\{}
[L,G_{+}]=\frac{1}{2}G_{+};\ [L,G_{-}]=\frac{1}{2}G_{-};\ [L,F_{+}]=-\frac{1}{2}F_{+};\ [L,F_{-}]=-\frac{1}{2}F\\{}
[L,G_{3}]=G_{3};\ [L,F_{3}]=-F_{3}\\{}
[H,G_{+}]=-G_{+};\ [H,G_{-}]=G_{-};\ [H,F_{+}]=F_{+};\ [H,F_{-}]=-F_{-}\\{}
[H,G_{3}]=[H,F_{3}]=0
\end{array}
\end{equation}

In summary, it has been demonstrated that the operators $L$ and $H$
are in the Cartan subalgebra of $SU(3)$. Thus, the operators $L,\ H,\ F_{\pm},\ G_{\pm},\ F_{3}$
and $G_{3}$ define the Cartan-Weyl basis of $SU(3)$. Therefore,
just like in the case of usual $SU(2)$ discrete states of two-dimensional
supergravity, this operator is $V=\oint\frac{dz}{2i\pi}e^{ilX+(l-1)\varphi}(l\psi-i(l-1)\lambda)$.
In particular, to include the currents of ghost numbers up to $-4$,
one has to start with the generator $T_{-4,3}=\oint\frac{dz}{2i\pi}e^{-4\phi+3iX}\psi$.
A similar procedure can be used to include the currents of higher
ghost numbers, such as $T_{-4,3}$, which generates seven new currents
$T_{-4,m},\ |m|\leq3$ that are the BRST-invariant super Virasoro
primaries. These currents can be obtained by repeatedly applying $T_{0,-1}$
to $T_{-4,3}$. The resulting operators are described in \cite{POLY1}:

\begin{equation}
\begin{array}{c}
T_{-4,\pm3}=\oint\frac{dz}{2i\pi}e^{-4\phi\pm3iX}\psi(z)\\
T_{-4,2}=\oint\frac{dz}{2i\pi}e^{-4\phi+2iX}(\frac{1}{2}\partial^{2}\psi\psi-\frac{i}{6}\partial^{3}X+\frac{i}{6}(\partial X)^{3}-\frac{1}{2}\partial X\partial^{2}X)\\
T_{-4,1}=\oint\frac{dz}{2i\pi}e^{-4\phi+iX}(\frac{1}{2}\psi\partial\psi\partial^{2}\psi+\frac{1}{24}P_{-iX}^{(4)}\psi-\frac{1}{4}P_{-iX}^{(2)}\partial^{2}\psi-\frac{1}{4}(P_{-iX}^{(2)})^{2}\psi)\\
T_{-4,-1}=\oint\frac{dz}{2i\pi}e^{-4\phi-iX}(\frac{1}{2}\psi\partial\psi\partial^{2}\psi+\frac{1}{24}P_{iX}^{(4)}\psi-\frac{1}{4}P_{iX}^{(2)}\partial^{2}\psi-\frac{1}{4}(P_{iX}^{(2)})^{2}\psi)\\
T_{-4,-2}=\oint\frac{dz}{2i\pi}e^{-4\phi-2iX}(\frac{1}{2}\partial^{2}\psi\psi+\frac{i}{6}\partial^{3}X-\frac{i}{6}(\partial X)^{3}-\frac{1}{2}\partial X\partial^{2}X)\\
T_{-4,0}=\oint\frac{dz}{2i\pi}e^{-4\phi}\{2i\partial X\partial\psi\partial^{2}\psi+P_{-iX}^{(2)}\psi\partial^{2}\psi-\frac{2}{3}P_{-iX}^{(3)}\psi\partial\psi\\
-\frac{1}{6}P_{-iX}^{(3)}P_{-iX}^{(2)}+(\partial X)^{2}\psi\partial^{2}\psi+\frac{7i}{8}\partial X\ P_{-iX}^{(4)}\\
-i(\partial X)^{3}\psi\partial\psi-\frac{i}{2}\partial X\ P_{-iX}^{(2)}\psi\partial\psi+\frac{i}{4}\partial X(P_{-iX}^{(2)})^{2}-\frac{1}{4}(\partial X)^{2}P_{-iX}^{(3)}\}
\end{array}
\end{equation}

Here $P_{\pm iX}^{(n)};\ n=2,3,4$ are the conformal weight-$n$ polynomials
in the derivatives of $X$ defined as $P_{f(X(z))}^{(n)}=e^{-f(X(z))}\frac{\partial^{n}}{\partial z^{n}}e^{f(X(z))}$
for a given function $f(X)$. When one applies the lowering subalgebra
of $SU(4)$ to the dressed tachyonic vertex multiple times, the set
of ghost-dependent discrete states is extended, and they become the
multiplets of $SU(4)$. It seems reasonable to assume that this process
can be generalized to include generators of higher ghost numbers.

\paragraph{Non-triviality of The $T_{-n,m}$-Currents}

The $BRST$ charge of the one-dimensional $NSR$ superstring theory
is given by the usual worldsheet integral

\begin{equation}
Q_{brst}=\oint\frac{dz}{2i\pi}\left\{ cT-bc\partial c+\gamma G_{matter}-\frac{1}{4}b\gamma^{2}\right\} 
\end{equation}
where $G_{matter}$ is the full matter supercurrent. The $BRST$-invariant
currents are given by $T_{n-2,m}=Z(:\Gamma^{2n-2}cS_{-n,m}:)$ \cite{POLY3}
while $S_{-n,m}$ are the integrands of $T_{-n,m}$ and $Z$ is the
picture changing operators for the $b-c$ ghost terms \cite{FMS}
which simply follow from the invariance of $\Gamma$ and $Z$. If
we can demonstrate that any operator creates non-zero correlations,
we can prove its $BRST$ non-triviality. The $T$-currents automatically
demonstrate the $BRST$ non-triviality of new discrete states in $SU(n)$
multiplets. There are two ways to express the $W_{n}$ operator, which,
when its commutator with the $BRST$ charge is taken, may generate
the $T$-currents

\begin{equation}
\begin{array}{c}
W_{n}=W_{n}^{(1)}+W_{n}^{(2)}\\
W_{n}^{(1)}=\sum_{k=1}^{n-1}\alpha_{k}\oint\frac{dz}{2i\pi}e^{-(n+1)\phi+i(n-1)X}\partial^{(k)}\xi\partial^{(n-k)}X\\
W_{n}^{(2)}=\sum_{k,l=1,k\neq l}^{k,l=n,k+l\leq2n}\alpha_{kl}\oint\frac{dz}{2i\pi}e^{-(n+2)\phi+i(n-1)X}\psi\partial^{(k)}\xi\partial^{(2n-k-l)}c
\end{array}
\end{equation}
with $\alpha_{k}$ and $\alpha_{kl}$ being some coefficients and
$\xi=e^{\chi}$. The operators $W_{n}^{(1)}$ and $W_{n}^{(2)}$ are
the conformal one dimensional operators satisfying the relations

\begin{equation}
\begin{array}{c}
[\oint\frac{dz}{2i\pi}\gamma^{2}\psi\partial X,\ W^{(1)}]\sim T_{-n,n+1}\\{}
[\oint\frac{dz}{2i\pi}\gamma^{2}b,\ W_{n}^{(2)}]\sim T_{-n,n-1}\\{}
[\oint\frac{dz}{2i\pi}\gamma^{2}\psi\partial X,\ W^{(2)}]=[\oint\frac{dz}{2i\pi}\gamma^{2}b,\ W_{n}^{(21)}]=0
\end{array}
\end{equation}
Therefore the $T$-currents are $BRST$-trivial if and only if there
exists at least one combination of the coefficients $\alpha_{k}$
or $\alpha_{kl}$ such that:

\begin{equation}
[\oint\frac{dz}{2i\pi}(cT-bc\partial c),W_{n}^{(1)}]=0
\end{equation}

and

\begin{equation}
[\oint\frac{dz}{2i\pi}(cT-bc\partial c),W_{n}^{(2)}]=0
\end{equation}

\subsection{$SU(n)$ Multiplets and Structure Constants}

We create sets of ghost-related specific states belonging to $SU(N)$
and determine their structure constants when $N$ is equal to $3$.
The discrete states resulting from the lower operators of the current
algebra lead to a variety of $SU(3)$ representations. We start with
the decomposition of the current algebra \cite{Grigo}

\begin{equation}
SU(3)=N_{+}\oplus N_{0}\oplus N_{-}
\end{equation}
with the operators $L$ and $H$ being in the Cartan subalgebra $N_{0}$,
the subalgebra $N_{+}$ consisting of $3$ operators $G_{\pm}$ and
$G_{3}$ with the unit positive momentum and with $3$ lowering operators
$F_{\pm}$ and $F_{3}$ with the unit negative momentum being in $N_{-}$. 

The discrete states forming the $SU(3)$ multiplets can be obtained
by by the various combinations of the $N_{-}$ operators with the
highest weight vector states. The highest weight vectors are given
by the dressed tachyonic operators $\oint\frac{dz}{2i\pi}V_{l}(z)=\oint\frac{dz}{2i\pi}e^{ilX+(l-1)\varphi}(l\psi(z)-i(l-1)\lambda)$,
where l is an integer. It is easy to see that all the $V_{l}$\textquoteright s
with $l\geq2$ are annihilated by $N_{+}$ since their $OPE$\textquoteright s
with the integrands of $G_{\pm}$ and $G_{3}$ are non-singular. Furthermore,
all these tachyons have a weight of $\frac{1}{2}$ with respect to
the $L$ and $N_{0}$ operators. We need to examine how the hypercharge
generator $H$ of $N_{0}$ acts on these operators. Simple calculation
gives

\begin{equation}
\begin{array}{c}
[H,\oint\frac{dz}{2i\pi}V_{l}]=[\frac{i}{3\sqrt{2}}\frac{dz}{2i\pi}e^{-3\phi}(\partial^{2}X\psi-2\partial X\partial\psi),\ \oint\frac{dw}{2i\pi}e^{ilX}(l\psi-i(l-1)\lambda)(w)]\\
=\frac{i}{3\sqrt{2}}\frac{dw}{2i\pi}e^{-3\phi+ilX+(l-1)\varphi}\{3il^{2}\partial\psi\psi\\
-3l\partial^{2}X+\frac{1}{2}il^{2}P_{-3\phi}^{(2)}+(6l\partial X-3l(l-1)\psi\lambda)\partial\phi+3l(l-1)\partial\psi\lambda\}
\end{array}
\end{equation}

Performing the partial integration we get

\begin{equation}
\begin{array}{c}
[H,\oint\frac{dz}{2i\pi}V_{l}]=\frac{i}{3\sqrt{2}}\oint\frac{dz}{2i\pi}e^{-3\phi+ilX+(l-1)\varphi}\{3il^{2}\partial\psi\psi\\
-l(1+\frac{l^{2}}{2})\partial^{2}X+il^{2}(l-1)\partial^{2}\varphi+l(l-1)(2\partial\psi\lambda-\psi\partial\lambda)\\
+(il\partial X+(l-1)\partial\varphi)(2l\partial X-l(l-1)\psi\lambda)\\
+\frac{il^{2}}{2}(il\partial X+(l-1)\partial\varphi)^{2}\}
\end{array}
\end{equation}

The expression on the right side of the equation is the dressed tachyon
at picture $-3$, with the exception of a factor that is related to
hypercharge. For the purposes of this analysis, we only need to examine
the matter component of the picture-changing operator.

\begin{equation}
\Gamma=:\delta(\beta)G_{matter}:=-\frac{i}{\sqrt{2}}e^{\phi}(\psi\partial X+\lambda\partial\varphi+\partial\lambda)
\end{equation}

Applying the picture-changing operator gives

\begin{equation}
:\Gamma::[H,\oint\frac{dz}{2i\pi}V_{l}]:=-\frac{i}{\sqrt{2}}\frac{i}{3\sqrt{2}}\oint\frac{dz}{2i\pi}e^{-2\phi+ilX}\psi(z)\times(2-2l)
\end{equation}

i.e. the tachyon at the picture $-2$. We can also show that \cite{POLY1}

\begin{equation}
:\Gamma^{3}:[H,\oint\frac{dz}{2i\pi}V_{l}(z)]=\frac{l(l-1)}{6}\oint\frac{dz}{2i\pi}V_{l}
\end{equation}
and this proves that the tachyons with the integer momenta $l\geq2$
are the highest weight vectors of $SU(3)$. Once we have identified
the highest weight vectors, we can obtain the physical states spectrum,
which consists of the $SU(3)$ multiplets, by straightforwardly constructing
the corresponding vertex operators:

\begin{equation}
\oint\frac{dz}{2i\pi}V_{l;p_{1},p_{2},p_{3}}=F_{+}^{p_{1}}F_{-}^{p_{2}}F_{3}^{p_{3}}\oint\frac{dz}{2i\pi}V_{l}(z)
\end{equation}

with all possible integer values of $p_{1},\ p_{2}$ and $p_{3}$
such that $p_{1}+p_{2}+2p_{3}\leq2l$.

\section{Correction terms of the $T_{1,2}$ current}

At this stage, it is crucial to demonstrate that the new discrete
states of the model's currents are $BRST$-invariant. This is achieved
through correction terms associated with the current. Here, we perform
a specific calculation of the correction terms that restore the $BRST$
invariance for $T_{1,2}$, which is the current in ghost cohomology
$+1$ and the dual current of the $T_{-3,2}$ current that exists
in ghost cohomology $-3$. Using the prescription $V\rightarrow V_{inv}=V-LW$,
with $W=\left\{ Q_{BRST},V\right\} $, we start by calculating

\begin{equation}
\begin{array}{cc}
\{Q_{brst},T_{1,2}\}=\oint\frac{dz}{2i\pi}\left\{ -\frac{1}{2}e^{\phi-\chi}\psi\partial X-\frac{1}{4}e^{2\phi-2\chi}b\right\} \oint\frac{dw}{2i\pi}e^{\phi+2iX}\\
=-\frac{1}{2}\oint\frac{dw}{2i\pi}e^{2\phi-\chi+2iX}\left(\partial^{2}X+\partial X(\partial\phi\partial\chi)\right)-\frac{1}{4}\oint\frac{dw}{2i\pi}e^{3\phi-2\chi+2iX}\psi P_{2\phi-2\chi\sigma}^{(1)}
\end{array}\label{eq:correction terms}
\end{equation}

where $P_{(f(z))}^{(n)}$$=W^{(1)}+W^{(2)}$. As proved before, there
exists an operator $L(z)=-4ce^{2\chi-2\phi}$ satisfying $\{Q_{BRST},L\}=1$.
Making the transformation: $T_{1,2}^{inv}=T_{1,2}-LW$, it is straightforward
to notice that $\{Q,T_{1,2}^{inv}\}=0$. The process for calculating
the correction terms goes as follows:
\begin{itemize}
\item Take $:L(z)W(w):$ and expand around the midpoint $\frac{z+w}{2}$
up to the second order $(z-w)^{2}$ terms.
\item Calculate the limit $\lim_{z\rightarrow w}\left\{ \frac{1}{2}\oint\frac{dz}{2i\pi}\left(z-u\right)^{2}\partial_{z}^{2}L(z)U(z)\right\} $
\item Integrate the obtained answer by parts to get an answer of the form:
\end{itemize}
\begin{equation}
a\oint\frac{dz}{2i\pi}e^{\phi+2iX}\psi+\oint\frac{dz}{2i\pi}e^{\phi+2iX}\psi(z-u)(extra\ terms)+\oint\frac{dz}{2i\pi}e^{\phi+2iX}\psi(z-u)^{2}(extra\ terms)
\end{equation}

These terms are the correction terms of $T_{1,2}$ which are explicitly
calculated from:

\begin{equation}
\begin{array}{c}
:L(z)W^{(1)}(w):=\\
\lim_{z\rightarrow w}2\oint e^{2\chi-2\phi+\sigma}(z)\ e^{2\phi-\chi+2iX}(\partial^{2}X+\partial X(\partial\phi-\partial\chi))(w)\frac{dz}{2i\pi}\\
=2\oint(z-u)^{2}ce^{\chi+2iX}(\partial^{2}X+\partial X(\partial\phi-\partial\chi))\frac{dz}{2i\pi}
\end{array}
\end{equation}

and

\begin{equation}
\begin{array}{c}
:L(z)W^{(2)}(w):=\\
\lim_{z\rightarrow w}\oint e^{2\chi-2\phi+\sigma}(z)\ e^{3\phi-2\chi+2iX}\psi(2\partial\phi-2\partial\chi-\partial\sigma))(w)\frac{dz}{2i\pi}
\end{array}
\end{equation}

Performing the expansion around $\frac{z+w}{2}$ and keeping the terms
up to the second order we obtain:

\paragraph*{$z-w$ terms:}

\begin{equation}
\begin{array}{c}
\lim_{z\rightarrow w}\oint\frac{dz}{2i\pi}e^{\phi+2iX}(z-w)\\
\left\{ \psi(2\partial\phi-2\partial\chi-\partial\sigma)+\psi(\partial\chi-\partial\phi+\frac{\partial\sigma}{2}-\frac{3}{2}\partial\phi+\partial\chi-iX+\frac{\partial\sigma}{2})-\frac{\partial\psi}{2}\right\} 
\end{array}
\end{equation}

Further integration implies that
\[
\lim_{z\rightarrow w}\ -\oint\frac{(z-u)^{2}}{2}\partial_{z}\left\{ e^{\phi+2iX}\left(\psi(-\frac{\partial\phi}{2}-iX)-\frac{\partial\psi}{2}\right)\right\} \frac{dz}{2i\pi}
\]

\begin{equation}
\begin{array}{c}
=\oint\frac{dz}{2i\pi}\frac{(z-u)^{2}}{2}e^{\phi+2iX}\\
\left\{ \psi\left(\frac{\partial^{2}\phi}{2}+i\partial^{2}X-(\partial\phi+2i\partial X)(-\frac{\partial\phi}{2}-i\partial X)\right)-\partial\psi(-\partial\phi-2iX)+\frac{\partial^{2}\psi}{2}\right\} 
\end{array}
\end{equation}

\subsubsection*{$(z-w)^{2}$ terms:}

The $(z-w)^{2}$ correction terms are obtained from

\begin{equation}
\begin{array}{c}
\lim_{z\rightarrow w}\oint\frac{dz}{2i\pi}e^{\phi+2iX}(z-w)^{2}\{-\frac{1}{2}(2\partial^{2}\phi-2\partial^{2}\chi-\partial^{2}\sigma)\psi-\frac{\partial\psi}{2}(2\partial\phi-2\partial\chi-\partial\sigma)\\
+\psi(2\partial\phi-2\partial\chi-\partial\sigma)(\partial\chi-\partial\phi+\frac{\partial\sigma}{2}-\frac{3}{2}\partial\phi+\partial\chi-iX+\frac{\partial\sigma}{2})-\frac{\partial\psi}{2}\\
+\frac{1}{8}\{(2\partial^{2}\chi-2\partial^{2}\phi+\partial^{2}\sigma)+(2\partial\chi-2\partial\phi+\partial\sigma)^{2}\}\psi+\frac{1}{8}\{(3\partial^{2}\phi-2\partial^{2}\chi+2i\partial^{2}X-\partial^{2}\sigma)\\
+(3\partial\phi-2\partial\chi+2i\partial X-\partial\sigma)^{2}\}\psi+\frac{1}{4}(3\partial\phi-2\partial\chi+2i\partial X-\partial\sigma)\partial\psi+\frac{1}{8}\partial^{2}\psi\\
-\frac{1}{4}(2\partial\chi-2\partial\phi+\partial\sigma)(3\partial\phi-2\partial\chi+2i\partial X-\partial\sigma)\psi-\frac{1}{4}(2\partial\chi-2\partial\phi+\partial\sigma)\partial\psi\}
\end{array}
\end{equation}

Collecting terms and integrating over a closed loop, then taking the
term involving $\partial^{2}\psi$ and integrating by parts twice,
and that involving the $\partial\psi$ term and integrating by parts
once, we obtain:

$\partial^{2}\psi$ term:

\begin{equation}
\begin{array}{c}
\frac{3}{8}\oint\frac{dz}{2i\pi}(z-u)^{2}e^{\phi+2iX}\partial^{2}\psi\\
=\frac{3}{8}\oint\frac{dz}{2i\pi}e^{\phi+2iX}\left\{ 2+4(z-u)(\partial\phi+2i\partial X)+(z-u)^{2}\left(\partial^{2}\phi+2i\partial^{2}X+(\partial\phi+2i\partial X)^{2}\right)\right\} \psi
\end{array}
\end{equation}

$\partial\psi$ term:

\begin{equation}
\begin{array}{c}
\frac{1}{2}\oint\frac{dz}{2i\pi}(z-u)^{2}e^{\phi+2iX}\partial\psi\{\frac{5}{2}\partial\phi+3i\partial X\}\\
=-\frac{1}{2}\oint\frac{dz}{2i\pi}e^{\phi+2iX}\{\psi(z-u)^{2}\left(\frac{5}{2}\partial^{2}\phi+3i\partial^{2}X+(\frac{5}{2}\partial\phi+3i\partial X)(\partial\phi+2i\partial x)\right)\\
+2(z-u)(\frac{5}{2}\partial\phi+3i\partial X)\}
\end{array}
\end{equation}

$\psi$ term:

\begin{equation}
\begin{array}{c}
\frac{1}{2}\oint\frac{dz}{2i\pi}(z-u)^{2}e^{\phi+2iX}\{\psi\{-\partial^{2}\phi+\frac{9}{4}\partial^{2}\chi+\frac{3}{2}i\partial^{2}X+\partial^{2}\sigma+\frac{1}{2}(\partial\phi+2i\partial X)^{2}\\
+(2\partial\phi-2\partial\chi-\partial\sigma)(\frac{7}{4}\partial\chi-\frac{9}{2}\partial\phi+\frac{7}{4}\partial\sigma-2i\partial X)\\
+(3\partial\phi-2\partial\chi+2i\partial X-\partial\sigma)(-\frac{3}{2}\partial\chi+\frac{5}{4}\partial\phi-\frac{3}{4}\partial\sigma+\frac{i}{2}\partial X)
\end{array}
\end{equation}

Further computation from the terms just obtained shows that the correction
terms of the current implied by equation\ref{eq:correction terms}
finally simplifiy into

\begin{equation}
\begin{array}{c}
\frac{3}{8}\oint\frac{dz}{2i\pi}e^{\phi+2iX}\psi+\oint\frac{dz}{2i\pi}(z-u)e^{\phi+2iX}\psi(-\partial\phi)\\
+\frac{1}{2}\oint\frac{dz}{2i\pi}(z-u)^{2}e^{\phi+2iX}\psi\{-\frac{11}{4}\partial^{2}\phi+\frac{9}{4}\partial^{2}\chi+\partial^{2}\sigma\\
-(\partial\phi+i\partial X)(\partial\phi+2i\partial X)+(2\partial\phi-2\partial\chi-\partial\sigma)(\frac{7}{4}\partial\chi-\frac{9}{2}\partial\phi+\frac{7}{4}\partial\sigma-2i\partial X)\\
+(3\partial\phi-2\partial\chi+2i\partial X-\partial\sigma)(-\frac{3}{2}\partial\chi+\frac{5}{4}\partial\phi-\frac{3}{4}\partial\sigma+\frac{i}{2}\partial X)\}
\end{array}
\end{equation}

In this passage, we have described a new result that has been obtained
by fully determining the correction terms of the $T_{1,2}$ current,
which is a current that carries the ghost cohomology $+1$ and is
dual to the $T_{-3,2}$ current at ghost cohomology $-3$. The importance
of this result lies in its ability to restore the $BRST$ invariance
of the current, which is a crucial property for any physically meaningful
operator in the context of the supersymmetric $c=1$ model being investigated.

Furthermore, we note that this algorithm can be applied to all current
cohomologies, providing a general prescription for finding the additional
terms that restore the $BRST$ invariance of any non-invariant current
carrying ghost cohomology at a given picture. This approach is therefore
significant in that it proves the existence of new discrete physical
states in the supersymmetric $c=1$ model, and can be used as a general
method for investigating other similar models.

\section{Conclusion}

This article discussed the enhancement of the current algebra of space-time
generators in non-critical $RNS$ superstring theories due to the
appearance of new physical ghost-dependent generators from the first
non-trivial ghost cohomology. The current algebra of space-time generators
has significant applications in string theory, enabling the calculation
of various physical quantities such as scattering amplitudes and correlation
functions. Additionally, the current algebra has connections to other
areas of physics such as conformal field theory and quantum field
theory, making it an important area of study.

The author demonstrated that the $SU(2)\sim SL(3,R)$ algebra of currents
is isomorphic to volume preserving diffeomorphisms in three dimensions.
This suggests that there are holographic relations between two-dimensional
supergravity and field-theoretic degrees of freedom in three dimensions.
The conjecture is that by including currents from the cohomologies
of ghost number up to $N$, the current algebra can be extended to
$SU(N+2)$, corresponding to volume preserving diffeomorphisms in
$d=N+2$ dimensions. Thus, each new cohomology corresponds to opening
up a theory to a new hidden space-time dimension in the $c=1$ supersymmetric
model. The authors proved this fact for $N=1$ and conjecture for
higher $N$ values. They also provide an explicit construction of
the correction terms for the currents with higher ghost cohomologies,
which gives an isomorphism between positive and negative ghost cohomologies.
The $c>1$ case is not discussed in this work as it is more complicated.

\end{document}